%% file: template.tex
\documentclass{Interspeech}

\interspeechcameraready

\title{A Comparative Study on Proactive and Passive Detection \\ of Deepfake Speech}

\author[affiliation={1,2}]{Chia-Hua}{Wu}
\author[affiliation={1}]{Wanying}{Ge}
\author[affiliation={1}]{Xin}{Wang}
\author[affiliation={1}]{Junichi}{Yamagishi}
\author[affiliation={2}]{Yu}{Tsao}
\author[affiliation={2}]{Hsin-Min}{Wang}

\affiliation{}{National Institute of Informatics}{Japan}
\affiliation{}{Academia Sinica}{Taiwan}
\email{maxwu@iis.sinica.edu.tw, jyamagis@nii.ac.jp}

\usepackage{comment}
\usepackage{booktabs}
\usepackage{multirow}
\usepackage{graphicx}  % for \resizebox
\usepackage{xcolor}    % for cell coloring
\usepackage{colortbl}  % for table shading
\usepackage{lipsum}
\usepackage{subfig}
\newcommand{\wanying}[1]{\textcolor{teal}{#1}}
\usepackage{cite}

\newcommand{\bitp}{\texttt{1}}
\newcommand{\bitn}{\texttt{0}}
\newcommand{\tagreal}{\texttt{real}}
\newcommand{\tagfake}{\texttt{fake}}

\begin{document}

\maketitle
\begin{abstract}
Solutions for defending against deepfake speech fall into two categories: proactive watermarking models and passive conventional deepfake detectors. While both address common threats, their differences in training, optimization, and evaluation prevent a unified protocol for joint evaluation and selecting the best solutions for different cases. This work proposes a framework to evaluate both model types in deepfake speech detection. To ensure fair comparison and minimize discrepancies, all models were trained and tested on common datasets, with performance evaluated using a shared metric. We also analyze their robustness against various adversarial attacks, showing that different models exhibit distinct vulnerabilities to different speech attribute distortions. Our training and evaluation code is available at Github~\footnote{https://github.com/nii-yamagishilab/antispoofing-watermark}.
\end{abstract}

\section{Introduction}
\label{sec:intro}

Protection against artificially generated or manipulated deepfake speech can be categorized into passive defense and proactive defense.
Passive defense involves detecting whether speech from a content provider was generated by artificial intelligence, without any pre-processing of the input speech or prior assumptions about the generator. 
This approach mainly utilizes deep learning-based deepfake detectors~\cite{jung2022_aasist,zhu2024_slim} trained on datasets containing real and fake speech~\cite{wang2020_asvspoof2019,xie2025_codecfake}. 
In contrast, proactive defense (typically watermarking models~\cite{liu2024_timbre,roman2024_audioseal,chen2023_wavmark}) embeds a traceable but perceptually inaudible message into the speech before distributing it, allowing the speech recipient to extract the message and verify the authenticity of the source~\cite{liu2024_audiomarkbench}. 
% This robustness not only allows for the concealment of multi-bit information but also ensures that intermediaries cannot easily access or alter the embedded watermark~\cite{wang2024_speechrecovery}.

Despite differences in use cases and methodologies, both passive and proactive approaches are potential solutions to detect speech deepfake. 
It is hence meaningful to compare these two types of methods. 
%, especially for the search for effective deepfake detection strategies. 
However, there are few relevant studies, possibly due to the challenges of comparing them fairly and rigorously:
%\begin{itemize}
%\item 
1) Deepfake detectors and watermarking models are trained with different objectives and loss functions~\cite{zhang2021_oneclass,de2023encodec}, and their outputs vary in form and semantics. 
2) They are usually evaluated using different metrics~\cite{liu2024_audiomarkbench,kinnunen18b_dcf} tailored to their use case, making direct comparison difficult.
3) Their robustness is often tested with different 
% speech features, information 
%domains or signal regions
transmission conditions and manipulations  ~\cite{roman2024_audioseal,liu2024_timbre,kawa23_defense,tak2022_rawboost}.
4) There is no common database and protocol shared by the two research communities.
%It is unclear whether these factors are equally important in the two tasks.
%\end{itemize}

This work takes the initial step toward bridging the gap. 
We compare representative speech deepfake detectors and watermarking models for identifying speech deepfakes, using their original training recipes, but on common speech deepfake detection datasets and diverse transmission and manipulation conditions.  
Our results show that powerful passive deepfake detectors and watermarking models can achieve perfect (or near-perfect) results on the ASVspoof 2019 and 2021 logical access (LA) datasets, but suffer varying degrees of performance degradation when facing unseen transmissions and manipulations. A watermarking model called Timbre tends to be the most robust but is still vulnerable to certain codecs and pitch shift manipulation. Other passive and proactive models are more fragile.   

%We remind the readers that the watermarking models require access to the test set speech to be watermarked. We cannot rely on watermarking models if we cannot watermark the speech data before distributing it.   
%Also note that, some watermarking models need access to the generative models for collaborative training~\cite{juvela24-collaborative-watermarking, juvela2024audio, cheng2024hifi}, setting an even higher bar of requirement. Nevertheless, comparison with this type of watermarking models is left for future work.

%In the rest of this paper, we briefly review the two paradigms (\S~\ref{sec:recap}), explain how to compare them (\S~\ref{sec:method}), and present the experimental results (\S~\ref{sec:exp}). The conclusion is drawn in \S~\ref{sec:conclude}.

\section{Brief Review of Speech Deepfake Detection and Watermarking}
\label{sec:recap}

\subsection{Speech deepfake detection}
\label{sec:recap:deepfakedetect}

A speech deepfake detector 
%aims to identify artifacts left behind by speech generation algorithms. It 
acts as a binary classifier and decides whether the input waveform is \emph{real or fake}.\footnote{Fake data is referred to as presentation attack or \emph{spoofed} data when it is used to compromise automatic speaker verification (ASV) systems. In that context, human speech is called \emph{bona fide} data~\cite{iso/iecjtc1/sc37ISO2023}. We use the terms \emph{real} and \emph{fake} (or deepfake) to address more general cases. }
Given a waveform $x \in \mathcal{X}$, the detector functions as $ f_{d}: \mathcal{X} \to \mathcal{Y}$, where $\mathcal{Y}\triangleq\{\tagreal, \tagfake\}$ is the set of output labels. 
In most cases, the detector is composed of a scoring function $ g_{d}: \mathcal{X} \to \mathbb{R}$ and a decision logic $h_{d}: \mathbb{R}\to\mathcal{Y}$. The function $g_{d}$ produces a score $s\in\mathbb{R}$ indicating the likelihood that the input is $\tagreal$. After that, $h_{d}$ assigns a label $\hat{y}=\tagreal$ if $s$ is larger than a decision threshold $\tau_{d}$. 
In implementation, $g_{d}$ can be a deep neural network (DNN) with a sigmoid output function, and $s$ can be the output of the sigmoid or the log odds~\cite{bishopPattern2006} fed into the sigmoid.

The performance of deepfake detectors is often measured using equal error rate (EER)~\cite{iso/iecjtc1/sc37ISO2023,wu2015spoofing,todisco2017constant, kinnunen2017asvspoof,wang2020_asvspoof2019,liu_asvspoof_2023, jung2022_aasist, tak22_w2v_aasist}. 
Let a test sample set be $\{(x_i, y_i)\}_{i=1}^{N}$, where the $\tagreal$ and $\tagfake$ samples are indexed by $i\in\Lambda_{\tagreal}$ and $i\in\Lambda_{\tagfake}$, respectively.
After $g_{d}$ produces the scores $\{s_i\}_{i=1}^{N}$, the false acceptance rate of fake data $\widehat{P}_{\text{FA}}(\tau_{d})$ and the false rejection rate of real data $\widehat{P}_{\text{FR}}(\tau_{d})$ can be estimated by
%\begin{align}
%\widehat{P}_{\text{FA}}(\tau_{d}) = \frac{1}{|\Lambda_{\tagfake}|}\sum_{i\in\Lambda_{\tagfake}}\mathbb{I}({s_i > \tau_d}),\\
%\widehat{P}_{\text{FR}}(\tau_{d}) = \frac{1}{|\Lambda_{\tagreal}|}\sum_{i\in\Lambda_{\tagreal}}\mathbb{I}({s_i < \tau_d}),
%\widehat{P}_{\text{FA}}(\tau_{d}) = \sum_{i\in\Lambda_{\tagfake}}\frac{\mathbb{I}({s_i > \tau_d})}{|\Lambda_{\tagfake}|},\quad
%\widehat{P}_{\text{FR}}(\tau_{d}) = \sum_{i\in\Lambda_{\tagreal}}\frac{\mathbb{I}({s_i < \tau_d})}{|\Lambda_{\tagreal}|},
%\end{align}
$\widehat{P}_{\text{FA}}(\tau_{d}) = \sum_{i\in\Lambda_{\tagfake}}{\mathbb{I}({s_i > \tau_d})}/{|\Lambda_{\tagfake}|}$ and $
\widehat{P}_{\text{FR}}(\tau_{d}) = \sum_{i\in\Lambda_{\tagreal}}{\mathbb{I}({s_i < \tau_d})}/{|\Lambda_{\tagreal}|}$, where $\mathbb{I}$ is an indicator function.
The EER is then estimated by $\text{EER} \approx \frac{1}{2}\big(\widehat{P}_{\text{FA}}(\tau_{d}^*) + \widehat{P}_{\text{FR}}(\tau_{d}^*)\big)$, where $\tau_d^*=\arg\min_{\tau_{d}}|\widehat{P}_{\text{FR}}(\tau_{d})  - \widehat{P}_{\text{FA}}(\tau_{d})|$ is the threshold at which $\widehat{P}_{\text{FA}}(\tau_{d}^*)$ and $\widehat{P}_{\text{FR}}(\tau_{d}^*)$ are approximately equal.

An ideal speech deepfake detector should be able to generalize well to unseen deepfake data.
%, even without any prior knowledge.
%To generalize well to new or out-of-domain speech data, 
To achieve this, the detector must be trained using data from diverse deepfake generators and acoustic conditions. Strategies such as data augmentation using simulated artifacts~\cite{tak2022_rawboost,wang2023_copy_synthesis} and self-supervised learning (SSL)-based feature extraction ~\cite{tak22_w2v_aasist,zhu2024_slim,guragain2024_speech} are also useful.  

\subsection{Multi-bit watermarking}
\label{sec:review_watermark}
Watermarking usually requires two components: a watermark embedder, which inserts a multi-bit message into the carrier waveform, and a watermark detector, which extracts and reconstructs the message from the received waveform~\cite{ji2024_speechwatermark, singh24_silentcipher}. 

Let $\mathcal{M}=\{M_1, \cdots, M_K\}$ be a set of $K$ pre-defined watermark messages, where each message $M_n\in\{\bitp, \bitn\}^L$ has $L$ bits. 
We may use many ($K\gg2$) messages and assign them to different parties for ownership verification, or we can just set $K=2$ for deepfake detection (described in \S~\ref{sec:method}).
The watermark embedder acts as a function $f_e: \mathcal{X} \times \mathcal{M} \to \mathcal{X}'$, where $\mathcal{X}'$ represents the domain of the watermarked waveform $x'$. 
Given $x'$, the detector $f_w: \mathcal{X}' \to \mathcal{M}$ detects and recovers the watermark message. The detection can be done using $L$ binary classifiers, i.e., $f_w = (f_{w,1}, f_{w,2}, \cdots, f_{w,L})$, where the $l$-th classifier $f_{w,l}: \mathcal{X}'\rightarrow\{\bitp, \bitn\}$ decides whether the $l$-th bit is $\bitp$ or $\bitn$. Similar to the deepfake detector, $f_{w,l} = h_{w,l}\circ g_{w,l}$ can be decomposed into a scoring function $g_{w,l}$ and a decision function $h_{w,l}$, but the meaning of score $s_{w,l}$ produced by $g_{w,l}$ is different --- it indicates the likelihood of the $l$-th bit being $\bitp$.

An ideal watermarking model should accurately detect the embedded watermark message while ensuring that the watermark is imperceptible to human listeners~\cite{roman2024_audioseal}. 
%Therefore, evaluation metrics for watermarking models include not only the watermark detection accuracy, but also the perceptual quality of the watermarked waveform, signal-to-noise ratio (SNR), etc. 
%The watermark detection accuracy can be measured at the bit level as $\text{ACC}_{\text{bit}} = \frac{1}{NL}\sum_{i=1}^N\sum_{l=1}^{L}\mathbb{I}(\widehat{m}_{i,l} = m_{i,l})$, where $\widehat{m}_{i,l}$ and ${m}_{i,l}$ are the $l$-th bit of the detected and ground truth messages of the $i$-th test utterance. The accuracy can be measured at the utterance level, which requires that all the detected bits match the ground truth, i.e., $\text{ACC}_{\text{utt}} = \frac{1}{N}\sum_{i=1}^{N}\prod_{l=1}^L\mathbb{I}(\widehat{m}_{i,l} = m_{i,l})$. 
The accuracy of watermark detection can be measured at the bit level, i.e., the percentage of detected bits that match the ground truth. Note that most studies assume a pre-defined decision threshold when making the decision on each bit~\cite{liu2024_timbre,chen2023_wavmark,cheng2024hifi,wuAdversarial2023,o2024maskmark}.

Watermarking models need to embed watermark information in a way that is imperceptible to the human ear while allowing for its perfect extraction. Therefore, compared to passive deepfake detectors, watermarking models require more complex training schemes, some of which incorporate specific operations to forge or remove watermarks during training
%Compared to binary classification, training watermarking models requires a more complex training scheme. To enhance robustness against transmission noise and watermark forgery or removal attacks, distortion layers are often introduced after the message embedding process
~\cite{singh24_silentcipher, roman2024_audioseal, liu2024_timbre}.

\section{Evaluation of Speech Deepfake Detection and Watermarking}
\label{sec:method}

\begin{figure}[t]
    \centering
    % \flushleft
    \hspace*{-0.5cm}
    % {\includegraphics[width=1.04\linewidth]{image/flow chart (17).pdf}}
    {\resizebox{0.5\textwidth}{!}{\includegraphics{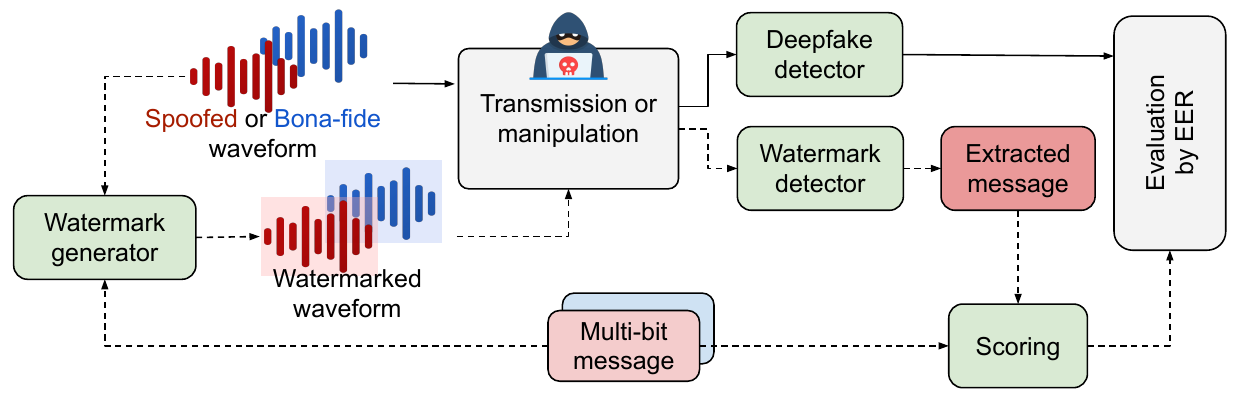}}}
    % \resizebox{\dimexpr\textwidth-1cm\relax}{!}{\includegraphics{image/flow chart (7).pdf}}
    % {\includegraphics[width=\linewidth, trim=0 300 390 0, clip]{image/figure.pdf}}

    % \vspace{-6mm}
    \caption{Overall comparison workflow between deepfake detection and watermarking. The solid arrows (\(\rightarrow\)) represent the deepfake detection process, while the dashed arrows (\(\dashrightarrow\)) indicate the watermarking process. %This diagram illustrates the key procedural differences between the two methods.
    }
    \vspace{-6mm}
    \label{fig:flow_chart}
\end{figure}

We now explain the evaluation framework shown in Fig.~\ref{fig:flow_chart}, upon which we compare deepfake detectors and watermarking models for binary deepfake detection.
%under diverse transmission conditions and manipulations. 
%A high-level illustration is shown in .

\subsection{Evaluation metrics}
For binary deepfake detection, we use EER as the main evaluation metric (\S~\ref{sec:recap:deepfakedetect}). EER is a concise summary of a detector's discriminative capabilities without pre-defined thresholds~\cite{van2007introduction}. It is also the upper-bound of Bayes decision error~\cite{brummer2021out}. 

%We also use the utterance-level balanced accuracy (Balanced Acc.) as a secondary metric~\cite{brodersen2010balanced},
%To account for potential data imbalance in test data and provide a more robust evaluation, we adopt balanced accuracy rather than simply reporting the highest observed accuracy \cite{brodersen2010balanced}. 
%Balanced accuracy is particularly useful when the class distribution is skewed because 
%which is computed as $1-\frac{1}{2}\Big(\widehat{P}_{\text{FR}}(\tau) + \widehat{P}_{\text{FA}}(\tau)\Big)$. %calculates the average recall rate, across the classes, thereby mitigating the bias caused by class imbalance. 
%The threshold $\tau$ is set at the point where the balanced accuracy is highest.
%This ensures a fairer assessment of model performance and prevents overestimation of accuracy in datasets with uneven class distributions.

The models to be evaluated, both deepfake detectors and proactive watermarking models, need to produce a continuous-valued detection score $s\in\mathbb{R}$ per test utterance (see \S~\ref{sec:recap:deepfakedetect}). The score implies how likely the utterance is $\tagreal$, and by convention~\cite{kinnunen18b_dcf}, we assume that a higher $s$ favors $\tagreal$ more. 

\subsection{Applying watermarking models to deepfake detection}
While deepfake detectors directly produce the score required, this is not the case for watermarking models.
Let us use two bit-wise disjoint \emph{random} messages $\mathcal{M}=\{M_\tagreal, M_\tagfake\}$ to watermark real and fake data, respectively. Each of the messages has $L$ bits, and each bit can be either $\bitp$ or $\bitn$.
To classify an input utterance, the watermark detector should act as $f_d: \mathcal{X}'\rightarrow \{\tagreal, \tagfake\}$, and its scoring function $g_d$ should produce the score $s$ that indicates the likelihood of the utterance being $\tagreal$.
Given $\{s_{w,l}\}_{l=1}^L$ produced by the bit-level scoring functions $\{g_{w,l}\}_{l=1}^L$ (see \S~\ref{sec:review_watermark}), 
we need to merge $\{s_{w,l}\}$ into $s$ but cannot simply let $s$ be the sum of $\{s_{w,l}\}$. Again, a higher $s_{w,l}$ suggests a more likely outcome of $\bitp$ but not necessarily $\tagreal$. 
%Whether $s_{w,l}$ favors $\tagreal$ depends on the $l$-th bit of $M_{\tagreal}$ (and $M_{\tagfake}$).

A general method to produce the score for deepfake detection without changing the watermarking model weights is 
\begin{equation}
    s =  \frac{1}{L}\sum_{l=1}^{L} \Big(s_{w,l} \cdot q(m_{\tagreal,l}) - s_{w,l} \cdot q(m_{\tagfake,l})\Big),
    \label{eq:score}
\end{equation}
where $m_{\tagreal,l}$ and $m_{\tagfake, l}$ are the $l$-th bit of the messages $M_{\tagreal}$ and $M_{\tagfake}$, respectively, and $q$ is a sign function that gives $q(\bitp)=1$ and $q(\bitn)=-1$. 

For explanation, let us use a simple example of $L=1$. If $m_{\tagreal, 1} = \bitp$ and $m_{\tagfake, 1} = \bitn$, we get $s = 2s_{w,1}$. A higher $s_{w,1}$, which favors $\bitp$ more, also gives more support to $\tagreal$. In the other case where $m_{\tagreal, 1} = \bitn$ and $m_{\tagfake, 1} = \bitp$, although a higher $s_{w,1}$ still favors $\bitp$ more, the score $s=-2s_{w,1}$ becomes smaller and gives less support to $\tagreal$ but more to $\tagfake$.\footnote{If $m_{\tagreal, l} =  m_{\tagfake, l}$, $s_{w,l}$ would give no support to either $\tagreal$ or $\tagfake$. This bit is hence a waste of the watermark capacity.}

\subsection{Evaluation with transmission and manipulation}
\label{sec:attack}
Both deepfake detection and watermark detection must perform robustly in clean environments as well as environments with noise, compression, encoding, etc. Both methods should also be robust to surreptitious waveform manipulations designed to flip detection results. This includes not only simple operations such as clipping and trimming but also advanced ones such as adding (or removing) background noise and using sound effects such as equalization and overdrive. 
%Both deepfake detection or watermarking models must be robust under these noisy transmission channels and, optionally, the acoustic manipulations. 

The transmission effects and manipulations used in the evaluation are listed below. 
Note that all conditions assume that the waveforms to be processed have a sampling rate of 16 kHz.

%It is important to evaluate the robustness of deepfake detectors and watermark models when the waveform is degraded in diverse transmission and manipulation conditions. We investigated the literature of both research fields and include 25 such conditions. We group them into two categories based on whether their mainly simulate the transmission effects or 
%In each attack, a set of randomly selected parameters will be adjusted. The following lists the attack methods and parameter details for the two groups:

\noindent\textbf{Group 1: Noisy and public transmission channels}
\begin{itemize}
\item \textbf{Gaussian noise}: Add a random Gaussian noise at a randomly selected SNR of 5, 10, or 15 dB.
\item \textbf{MUSAN}: Add a randomly selected noise from the MUSAN corpus~\cite{snyder_musan_2015} to the waveform at a fixed SNR of 10 dB.  
\item \textbf{RIR}: Convolve signals with a simulated room impulse response randomly selected from the RIR corpus~\cite{ko_study_2017}.
\item \textbf{Quantization}: Quantize the waveform using a randomly selected bit-depth of 8, 16, 24, or 32 to add quantization noise. 
\item \textbf{Compressor}: Apply dynamic range compression \cite{sobot_peter_2023_7817838} with a random threshold (-50 to -10 dB) and ratio (2.0 to 10.0).% using Pedalboard.
% \item \textbf{MP3}: compress the waveform using MP3 with a randomly selected quality level (0.0, 0.5, or 1.0) using Pedalboard\cite{sobot_peter_2023_7817838}.
\item \textbf{Opus}: Compress using the Opus codec~\cite{valin2012definition} with a random bitrate (1, 2, 4, 8, 16, or 31 kbps) to simulate VoIP/streaming.  
% \item \textbf{EnCodec}~\cite{defossez2022high}: compress using the 24 kHz Encodec model at a random bitrate (1.5, 3, 6, 12, or 24 kbps). 
\item \textbf{DAC}: Compress 16 kHz audio into discrete codes and then decompress it back into an audio signal~\cite{kumar2023high}.
\item \textbf{WavTokenizer}: Tokenize and reconstruct the waveform using WavTokenizer (small-600-24k-4096)~\cite{ji2024wavtokenizer}, a state-of-the-art neural audio codec. 
\end{itemize}
\noindent\textbf{Group 2: Acoustic manipulations}
\begin{itemize}
% \item \textbf{Amplification}: adjust the signal amplitude by a randomly selected gain factor between 1.1 and 2.0.  
\item \textbf{Clipping}: Apply amplitude clipping and constrain the amplitude to be within the 1st and 99th percentile range.  
\item \textbf{Overdrive}: Implement Overdrive using the PyTorch\cite{paszke2019pytorch} function, which introduces nonlinear distortion with a randomly selected gain (0 to 50 dB) and colour (0 to 50).  
\item \textbf{Random trimming}: Trim the waveform with randomly selected start and end times.
% \item \textbf{Smoothing}: perform local averaging on the waveform using a randomly selected window size (4, 8, or 16). %to reduce fluctuations.
% \item \textbf{Low-pass} or \textbf{High-pass Filtering}: apply low-pass or high-pass filtering with a random cutoff frequency between 1 and 5 kHz or between 20 Hz and 2.4 kHz.  
%\item \textbf{High-pass Filtering}: apply high-pass filtering with a  random cutoff frequency between 20Hz and 2.4kHz.
\item \textbf{Equalizer}: Apply a 7-band parametric equalizer with a random gain range between -12 and 12dB.
\item \textbf{Frequency masking}: Randomly zero out 10 to 80 Short-time Fourier Transform (STFT) frequency bins of the waveform and reconstruct the waveform via inverse STFT (iSTFT).
%apply frequency Masking by computing STFT, zeroing a random 10-80 frequency bin range, and converting it back with inverse STFT.
% \item \textbf{Upsampling} or \textbf{Dowmsampling}: randomly resample to 4, 8, 22.05, 24, 44.1, or 48 kHz. The waveform is then resampled back to 16 kHz.
%\item \textbf{Downsampling}: randomly resample to either 4 or 8 kHz.  
% \item \textbf{Noise-gate}: apply the noise-gate function of Audacity~\cite{audacityteamAudacityR2014} to allow waveform above a specified threshold level to pass
\item \textbf{Noise gate}: Compute a spectrogram, estimate noise thresholds, applying gating, and reconstructing the signal\cite{sainburg2020finding, tim_sainburg_2019_3243139}.
\item \textbf{Noise reduction}: Apply a DNN-based speech enhancement model~\cite{schroeter2022deepfilternet, schroeter2023deepfilternet3} to the waveform.
\item \textbf{Time stretch}: Use a phase vocoder via Librosa~\cite{mcfee2015librosa} to stretch the waveform with a random rate between 0.5 and 2.0.
\item \textbf{Pitch shift}: Use a phase vocoder and resampling via Librosa~\cite{mcfee2015librosa} to randomly shift the pitch between $\pm 5$ semitones.
\end{itemize}
%In each attack, a set of randomly selected parameters will be adjusted. 
%\wanying{In addition to evaluating the selected deepfake detectors and watermarking models using clean speech files, we also evaluate their robustness against degraded or manipulated waveforms by introducing artificial distortions. These distortions simulate real-world conditions, such as transmission channels affected by noise and compression, as well as malicious attacks intended to bypass detection by reversing the original decision. Secure and reliable protection in such environments requires models that can produce robust and (nearly) consistent scoring results, not only under simple signal processing operations like clipping, trimming, and encoding, but also under more aggressive acoustic manipulations, such as background noise and sound effects (e.g., EQ, overdrive, and those generated by neural networks).}
%The above transmission effects and manipulation are applied to the waveform fed into the deepfake detector and the watermark detector. 
%For deepfake detectors, the transmission effects and manipulations are applied to the waveform input. 
As shown in Fig.~\ref{fig:flow_chart}, for watermarking models, the above processing is applied to the watermarked waveform input to the watermark detector, rather than the waveform input to the watermark encoder.
%The inputs to the watermark encoder remain clean and unprocessed as Figure~\ref{fig:flow_chart} shows. 

We also investigated other transmission and manipulation types. However, for a fair comparison, they were excluded from the experiments because they were seen during the training of one or more of the compared models. The above transmissions and manipulations that are similar but not identical to those used in training are considered partially seen evaluation conditions. For example, if a DNN-based codec called EnCodec~\cite{de2023encodec} is seen in AudioSeal training, transmissions using DAC or WavTokenizer are considered partially seen because they are similar to EnCodec in terms of DNN architecture and training criteria.
%DNN-based codec similar to  Encodec~\cite{de2023encodec} used by AudioSeal for data augmentation, hence marked as partially seen. We summarize the findings as below.
%Readers who are interested in results on the seen transmission and manipulation types are encouraged to check the appendix.

\subsection{Dataset}
\label{sec:method:data}
The deepfake detection and watermarking communities use different databases for experiments. To compare the two methods for deepfake detection, we need a common database containing real and fake data and well-designed protocols.
%supporting the training and evaluation of both methods.

We follow the speech deepfake detection community and use the LA part of the ASVspoof 2019~\cite{wang2020_asvspoof2019} and ASVspoof 2021~\cite{liu_asvspoof_2023} datasets. 
%Both contain real and fake data as well as official protocols. 
Following the official protocol~\cite{liu_asvspoof_2023}, we utilize the training and development sets of the ASVspoof 2019 LA dataset for model training and validation. These datasets include fake data generated using 6 different algorithms (A01–A06 in \cite{wang2020_asvspoof2019}). 
The ASVspoof 2019 test set and the entire 2021 LA set are used for testing, comprising fake data from 11 unseen and 2 seen deepfake generators (A07-A19 in \cite{wang2020_asvspoof2019}). %The number of real and fake utterances for testing are around 14k and 133k, respectively. 
Compared to the 2019 test set, 
%Furthermore, except for one specific condition, all bona fide and spoofed speech samples underwent processing through seven different audio codecs during transmission, simulating real-world application scenarios.
around 85\% of data in the 2021 dataset has been pre-processed under lossy transmission channels (e.g., PSTN).\footnote{If a channel-transmitted utterance is further processed by any transmission in \S~\ref{sec:attack}, it is called double degradation~\cite{liu_asvspoof_2023}.} There is no speaker overlap in the training, development, and test sets. 
Other details about the data and protocols can be found in~\cite{wang2020_asvspoof2019,liu_asvspoof_2023}.

\input{table_result_4}

\section{Experiment}
\label{sec:exp}
To assess the proposed evaluation framework, we conducted a comparative study using four representative DNN-based deepfake detectors and watermarking models. 
%(\S~\ref{sec:exp:models}). The results are represented in \S~\ref{sec:exp:results}.

%To assess performance, we adopt two evaluation metrics: equal error rate (EER) and balanced accuracy (Balanced Acc).

%We report both metrics to provide a comprehensive evaluation. EER is the standard metric for deepfake detection, while accuracy is commonly used in audio watermarking. However, our accuracy metric differs from conventional audio watermarking accuracy, which is typically measured at the bit level and selects the highest correct rate. Instead, we evaluate accuracy at the sentence level to align with deepfake detection. Furthermore, to ensure a fairer comparison and mitigate the impact of data imbalance, we report balanced accuracy rather than highest accuracy.

\subsection{Experimental models} 
\label{sec:exp:models}

%\subsubsection{Deepfake detection} %
%\label{sec:deepfake_detection}
\textbf{Passive deepfake detectors}: 
The two deepfake detectors we studied are \textbf{AASIST}~\cite{jung2022_aasist} and \textbf{SSL-AASIST}~\cite{tak22_w2v_aasist}. 
{AASIST} is a state-of-the-art end-to-end (E2E) spoofing countermeasure solution. It uses a sinc-layer front-end to decompose raw waveforms for feature extraction and a graph attention network back-end to integrate temporal and spectral representations, followed by a readout operation and a fully connected output layer to produce decision scores.
{SSL-AASIST} enhances AASIST by integrating the pre-trained wav2vec2.0 model \cite{baevski2020wav2vec} as the front-end for feature extraction. %This model has been trained on 437,000 hours of bona fide speech from five large speech datasets, covering 128 languages and over 60,000 speakers. 
Using SSL models in the front end improves robustness against noise, reverberation, and other external distortions~\cite{tak22_w2v_aasist}. %, making SSL-AASIST more resilient in real-world conditions.
% We use official implementations and the released AASIST\footnote{AASIST: https://github.com/clovaai/aasist} and SSL-AASIST\footnote{SSL-AASIST: https://github.com/TakHemlata/SSL\_Anti\_spoofing} checkpoints, which were trained on the ASVspoof 2019 LA dataset. 
We use official implementations and the released AASIST and SSL-AASIST checkpoints, which were trained on the ASVspoof 2019 LA dataset. 
%Rather than retraining them, we directly apply the publicly available pre-trained models for our experiments.

% \noindent\textbf{Proactive watermarking models:} %
% The two experimental watermarking models are \textbf{Timbre}~\cite{liu2024_timbre}\footnote{Timbre: https://github.com/TimbreWatermarking/TimbreWatermarking} and \textbf{AudioSeal}~\cite{roman2024_audioseal}\footnote{AudioSeal: https://github.com/facebookresearch/audiocraft}. 
\noindent\textbf{Proactive watermarking models:} %
The two experimental watermarking models are \textbf{Timbre}~\cite{liu2024_timbre} and \textbf{AudioSeal}~\cite{roman2024_audioseal}. 
{Timbre} embeds and detects watermarks in the spectral domain. It applies STFT to extract the spectrogram, embeds the watermark in the amplitude while preserving the phase, and reconstructs the waveform using iSTFT. %This ensures robustness and minimal perceptual distortion.
%The original Timbre Watermarking model was trained on the LibriSpeech dataset. To fairly compare it with Deepfake detection models, we retrained it on the ASVspoof 2019 LA dataset using an open-source toolkit \footnote{https://github.com/TimbreWatermarking/TimbreWatermarking.git}. The sampling rate was standardized to 16,000 Hz, and the watermark message length was set to 16 bits.
AudioSeal is a state-of-the-art audio watermarking framework based on a sequence-to-sequence encoder-decoder architecture. 
%The encoder embeds watermarks while preserving perceptual quality, and the decoder extracts the watermark. %and provides a detection probability. A predefined threshold determines whether a watermark is present.

Following the official implementation, we trained the two proactive models using the ASVspoof 2019 LA training and development sets.
%The original AudioSeal model was trained on a 4.5K-hour subset of the VoxPopuli dataset \cite{wang2021voxpopuli}. To fairly compare it with Deepfake detection models, we retrained it on the ASVspoof 2019 LA dataset using Facebook's open-source Audiocraft toolkit \footnote{https://github.com/facebookresearch/audiocraft}. 
The sampling rate is 16 kHz, and the watermark message length is 16 bits. The outputs of Timbre's bit-level detectors' linear output layer and the sigmoid logit of the AudioSeal output layers are utilized as the score $s_{w,l}$ in Eq.~(\ref{eq:score}).
The official implementation of AudioSeal uses data augmentation based on MP3 compression, re-sampling, etc. As mentioned in \S~\ref{sec:attack}, transmissions and manipulations used in data augmentation for model training are excluded from the evaluation. 

% EER, AUC, DEC-Curve, Accuracy
% \begin{enumerate}
%     \item Data: ASVspoof 2019 \& 2021
%     \item Model training:
%     \begin{enumerate}
%         \item AASIST, Wav2Vec
%         \item Timbre, AudioSeal
%     \end{enumerate}
%     \item Metrics
%     \begin{enumerate}
%         \item EER
%         \item AUC, DEC-Curve
%         \item Accuracy

%     \end{enumerate}
%     \item Code link
% \end{enumerate}
%\subsection{Implementation details}

\subsection{Results and analysis}
\label{sec:exp:results}

Table~\ref{tab:performance} presents the results. Without the transmission or manipulation in \S~\ref{sec:attack}, the proactive models demonstrate the ability to perfectly distinguish real from spoofed speech, achieving 0\% EER on the ASVspoof 2019 LA and ASVspoof 2021 LA evaluation sets. This shows that watermarks can be accurately embedded into and extracted from real and fake utterances, even when they are uttered by an unseen speaker or produced by 11 unseen deepfake generators. 
Among the passive models, SSL-AASIST exhibits near-perfect performance, with EERs of 0.23\% and 0.84\% on the two test sets, respectively. However, while the EER on the 2019 LA test set is below 1\%, AASIST's EER on the 2021 LA data rises to 8.15\% because the data contains transmission channel operations. 

When the transmission or manipulation listed in \S~\ref{sec:attack} is applied, in many cases all model performance drops, even if the transmission or manipulation is similar to the data augmentation methods during training. 
%, it does not guarantee a low EER in this partially seen condition. 
For example, although AudioSeal has used EnCodec for data augmentation, similar codecs (DAC and wavTokenizer) result in EERs of 60.95\% and 97.40\% on the ASVspoof 2019 LA test set. An EER larger than 50\% means that watermarked fake utterances are more likely to be detected as \tagreal. Similarly, in the passive AASIST model, random trimming during training does not lead to a lower EER when random trimming is applied to test utterances. The only exception is SSL-AASIST under the Gaussian noise condition, probably due to the robustness of the pre-trained SSL model to simple additive noise.

Among the four models, Timbre appears to be the most robust, achieving EERs of 8.87\% and 9.02\% on two evaluation sets, followed by the passive model SSL-AASIST, which achieves 12.41\% and 14.34\%.  
However, we cannot hastily conclude that proactive models are more robust. First, AudioSeal's performance degrades severely when certain transmissions or manipulations are applied to watermarked waveforms. Second, even the best-performing Timbre model shows varying degrees of degradation when facing different transmission and manipulation types. For example, while Clipping and Equalizer do not harm EER, simple Gaussian noise results in an EER of 17.60\%. More complex techniques, such as DNN-based WavTokenizer and signal-processing-based Pitch shift, result in an EER of around 50\%, where the model is unable to differentiate real and fake data based on the detected watermarks.

%A notable observation is the discrepancy in performance between the two evaluation sets for passive models. Specifically, SSL-AASIST exhibits an EER increase of 1.93\%, while AASIST experiences a more pronounced increase of 5.9\%. 
%In contrast, proactive models maintain a relatively consistent performance across the two datasets, suggesting their greater resilience to channel effects. 
%This observation implies that proactive models may be more robust against variations introduced by different transmission and recording environments.

All the results suggest that robustness against transmission and manipulation is an unsolved issue. Challenging transmission and manipulation types are highlighted below. 
\begin{itemize}
%\item Noise Injection Attacks (Gaussian-Noise, MUSAN, RIR): Under noise injection attacks, SSL-AASIST demonstrates strong resilience, consistently achieving low EER values. Among the proactive models, Timbre ranks second overall but achieves the best performance under MUSAN and RIR attacks, with EERs of 1.31\% and 0.00\%, respectively. These results suggest that Timbre exhibits a high degree of adaptability to environmental noise perturbations.
\item Codecs (Opus, DAC, and WavTokenizer): 
While we have described how DAC and WavTokenizer induce high EER on AudioSeal, WavTokenizer also increases the EER of other proactive and passive models. %while DAC is less challenging.  
%In evaluating attacks related to audio compression and quantization, we observe that both SSL-AASIST and Timbre effectively mitigate these adversarial conditions, maintaining relatively low EER values. 
The non-DNN codec Opus also proved harmful to all models, especially AudioSeal. Different ways of reconstructing waveforms, namely linear prediction plus 
discrete cosine transform (Opus), transposed convolution (DAC), and convolution plus iSTFT (WavTokenizer), may affect how the watermark or deepfake artifact is altered.
%However, further analysis of attacks targeting codec (Opus) and neural codec (WavTokenizer) reveals that these approaches significantly degrade the performance of both passive and proactive models. 
%Notably, watermarking-based attacks appear to be particularly detrimental to proactive models, an issue that warrants further investigation.
\item Temporal and spectral modifications (Time stretch, Pitch shift, and Random trimming):
%These attacks involving temporal and spectral modifications introduce significant challenges for all models. 
While time-stretch mainly affects passive models, Pitch shift that re-samples time-stretched waveforms impairs all models. Random trimming will hurt the performance of models other than Timbre. This is expected because Timbre explicitly embeds the watermark in the frequency domain of each frame and is therefore robust to temporal manipulations. However, this design is vulnerable to frequency domain transformations.
%which will be discussed in greater detail in subsequent sections. 
%Overall, Timbre exhibits remarkable robustness against this category of attacks, with an EER close to 0.00\%, highlighting its superior adaptability to audio deformation. In contrast, the remaining models demonstrate less stability when exposed to such perturbations.
\end{itemize}

\vspace{-2mm}
\section{Conclusion}
\label{sec:conclude}
In this work, we took the initiative to compare representative passive deepfake detectors (AASIST and SSL-AASIST) and proactive watermarking models (Timbre and AudioSeal) for the binary classification task of speech deepfake detection. 
%While it appears to be straightforward, the comparison requires a unified evaluation metric, scoring method, diverse evaluation conditions, as well as a common dataset. 
Comparisons based on unified evaluation metrics, scoring methods, diverse evaluation conditions, and common datasets show that watermarking models and SSL-AASIST can perfectly detect speech deepfakes without transmission or manipulation, but all models fail to varying degrees under different transmission or manipulation conditions. 
Timbre appears to be the most robust, but it still suffers from pitch shift based manipulation or transmission via WavTokenizer or Opus. 
%The role of data augmentation in shaping model robustness cannot be overlooked, as it significantly influences the learned representations and generalization capabilities. Therefore, we cannot definitively conclude whether watermark-based detection or deepfake detection is superior. Instead, our study provides a preliminary comparative analysis based on the baseline configurations of each model, offering insights into their respective strengths and weaknesses. 

Robustness to various transmissions and manipulations must be better addressed for the passive and proactive models investigated. This initial study used the official training recipes, some of which did not incorporate any data augmentation. 
%Finally, given that the training objectives and data augmentation strategies of passive and proactive models differ, 
Future work will investigate the impact of data augmentations on proactive and passive models.

\vspace{-2mm}
\section{Acknowledgements}
This work was conducted during the first author’s internship at the National Institute of Informatics (NII), Japan. This study was partially supported by JST AIP Acceleration Research (JPMJCR24U3), MEXT KAKENHI Grant (24H00732), and JST PRESTO (JPMJPR23P9).

\bibliographystyle{IEEEtran}
\bibliography{mybibv4}

\end{document}

%% file: table_result_4.tex
\begin{table*}[ht]
    \centering
    \caption{Equal Error Rate (EER) results on the ASVspoof 2019 LA and ASVspoof 2021 LA datasets. Darker-shaded values (e.g., in gray) indicate worse performance, corresponding to higher EER values. A transmission or manipulation condition is considered as partially seen if it is used by any of the experimental models during training. The model partially seen the condition is marked with an asterisk $\ast$. The two sub-groups in the partially seen and unseen conditions correspond to transmission and manipulation, respectively.}
    \vspace{-2mm}
    \resizebox{\textwidth}{!}{
    \begin{tabular}{clp{1.6cm}p{1.8cm}p{1.6cm}p{1.6cm}p{1.6cm}p{1.8cm}p{1.6cm}p{1.6cm}}
        \toprule
        \multirow{3}{*}{\textbf{}} & \multirow{3}{*}{\shortstack{{\phantom{0}}\\ \underline{Transmission}\\Manipulation}} 
        & \multicolumn{4}{c}{EER (\%)$\downarrow$ of ASVspoof 2019 LA } 
        & \multicolumn{4}{c}{EER (\%)$\downarrow$ of ASVspoof 2021 LA} \\
        \cmidrule(lr){3-6} \cmidrule(lr){7-10}
        & & \multicolumn{2}{c}{Passive Models} & \multicolumn{2}{c}{Proactive Models}
        & \multicolumn{2}{c}{Passive Models} & \multicolumn{2}{c}{Proactive Models} \\
        \cmidrule(lr){3-4} \cmidrule(lr){5-6} \cmidrule(lr){7-8} \cmidrule(lr){9-10}
        & & AASIST & SSL-AASIST & Timbre & AudioSeal & AASIST & SSL-AASIST & Timbre & AudioSeal \\ 
        \midrule
        &  None from \S~\ref{sec:attack} 
        & \cellcolor[rgb]{1.00, 1.00, 1.00}\phantom{0}0.83 & \cellcolor[rgb]{1.00, 1.00, 1.00} \phantom{0}0.23  & \cellcolor[rgb]{1.00, 1.00, 1.00} \phantom{0}0.00 & \cellcolor[rgb]{1.00, 1.00, 1.00} \phantom{0}0.00 & \cellcolor[rgb]{0.98, 0.98, 0.98}\phantom{0}8.15 & \cellcolor[rgb]{1.00, 1.00, 1.00}\phantom{0}0.84 & \cellcolor[rgb]{1.00, 1.00, 1.00}\phantom{0}0.00 & \cellcolor[rgb]{1.00, 1.00, 1.00}\phantom{0}0.00 \\ 
        \midrule\multirow{6}{*}{\rotatebox{90}{\shortstack{Partially \\ seen}}}
         & Gaussian noise & \cellcolor[rgb]{0.96, 0.96, 0.96} 18.06 
         & \cellcolor[rgb]{1.00, 1.00, 1.00} \phantom{0}1.95 $\ast$ 
         & \cellcolor[rgb]{0.96, 0.96, 0.96} 17.60 
         & \cellcolor[rgb]{0.96, 0.96, 0.96} 15.83 $\ast$ 
         & \cellcolor[rgb]{0.94, 0.94, 0.94} 25.00 
         & \cellcolor[rgb]{0.99, 0.99, 0.99} \phantom{0}2.79 $\ast$
         & \cellcolor[rgb]{0.96, 0.96, 0.96} 18.73 
         & \cellcolor[rgb]{0.96, 0.96, 0.96} 16.04 $\ast$  \\ 
         & DAC & \cellcolor[rgb]{1.00, 1.00, 1.00} \phantom{0}1.66 
         & \cellcolor[rgb]{1.00, 1.00, 1.00} \phantom{0}0.27 
         & \cellcolor[rgb]{1.00, 1.00, 1.00} \phantom{0}0.01 
         & \cellcolor[rgb]{0.59, 0.59, 0.59} 97.40 $\ast$
         & \cellcolor[rgb]{0.98, 0.98, 0.98} \phantom{0}8.44 
         & \cellcolor[rgb]{1.00, 1.00, 1.00} \phantom{0}1.44 
         & \cellcolor[rgb]{1.00, 1.00, 1.00} \phantom{0}0.00 
         & \cellcolor[rgb]{0.59, 0.59, 0.59} 97.59 $\ast$  \\ 
         & WavTokenizer & \cellcolor[rgb]{0.96, 0.96, 0.96} 17.84 
         & \cellcolor[rgb]{0.96, 0.96, 0.96} 15.92 
         & \cellcolor[rgb]{0.85, 0.85, 0.85} 50.12 
         & \cellcolor[rgb]{0.80, 0.80, 0.80} 60.95 $\ast$
         & \cellcolor[rgb]{0.96, 0.96, 0.96} 17.31 
         & \cellcolor[rgb]{0.96, 0.96, 0.96} 16.31 
         & \cellcolor[rgb]{0.86, 0.86, 0.86} 46.32 
         & \cellcolor[rgb]{0.78, 0.78, 0.78} 63.80 $\ast$  \\ 
         \cmidrule{2-10}
         & Random trimming & \cellcolor[rgb]{0.95, 0.95, 0.95} 19.56 $\ast$ 
         & \cellcolor[rgb]{0.98, 0.98, 0.98} \phantom{0}8.15 
         & \cellcolor[rgb]{1.00, 1.00, 1.00} \phantom{0}0.00 
         & \cellcolor[rgb]{0.89, 0.89, 0.89} 37.50 
         & \cellcolor[rgb]{0.93, 0.93, 0.93} 26.86 $\ast$ 
         & \cellcolor[rgb]{0.97, 0.97, 0.97} 11.48 
         & \cellcolor[rgb]{1.00, 1.00, 1.00} \phantom{0}0.00 
         & \cellcolor[rgb]{0.90, 0.90, 0.90} 36.87  \\ 
         & Time stretch & \cellcolor[rgb]{0.77, 0.77, 0.77} 66.53 
         & \cellcolor[rgb]{0.87, 0.87, 0.87} 44.42 
         & \cellcolor[rgb]{1.00, 1.00, 1.00} \phantom{0}0.00 
         & \cellcolor[rgb]{1.00, 1.00, 1.00} \phantom{0}0.03 $\ast$
         & \cellcolor[rgb]{0.76, 0.76, 0.76} 68.08 
         & \cellcolor[rgb]{0.86, 0.86, 0.86} 47.56 
         & \cellcolor[rgb]{1.00, 1.00, 1.00} \phantom{0}0.00 
         & \cellcolor[rgb]{1.00, 1.00, 1.00} \phantom{0}0.05 $\ast$  \\ 
         & Pitch shift & \cellcolor[rgb]{0.77, 0.77, 0.77} 66.12 
         & \cellcolor[rgb]{0.85, 0.85, 0.85} 48.36 
         & \cellcolor[rgb]{0.83, 0.83, 0.83} 52.62 
         & \cellcolor[rgb]{0.86, 0.86, 0.86} 47.30 $\ast$ 
         & \cellcolor[rgb]{0.76, 0.76, 0.76} 68.67 
         & \cellcolor[rgb]{0.85, 0.85, 0.85} 48.84 
         & \cellcolor[rgb]{0.83, 0.83, 0.83} 53.14 
         & \cellcolor[rgb]{0.85, 0.85, 0.85} 50.11 $\ast$  \\ 
         \midrule\multirow{11}{*}{\rotatebox{90}{Unseen}}
         & MUSAN & \cellcolor[rgb]{0.96, 0.96, 0.96} 17.84 
         & \cellcolor[rgb]{1.00, 1.00, 1.00} \phantom{0}1.73 
         & \cellcolor[rgb]{1.00, 1.00, 1.00} \phantom{0}1.31 
         & \cellcolor[rgb]{0.99, 0.99, 0.99} \phantom{0}2.91 
         & \cellcolor[rgb]{0.94, 0.94, 0.94} 26.31 
         & \cellcolor[rgb]{0.99, 0.99, 0.99} \phantom{0}2.97 
         & \cellcolor[rgb]{1.00, 1.00, 1.00} \phantom{0}2.02 
         & \cellcolor[rgb]{0.99, 0.99, 0.99} \phantom{0}3.25  \\ 
         & RIR & \cellcolor[rgb]{0.90, 0.90, 0.90} 35.49 
         & \cellcolor[rgb]{0.99, 0.99, 0.99} \phantom{0}4.41 
         & \cellcolor[rgb]{1.00, 1.00, 1.00} \phantom{0}0.00 
         & \cellcolor[rgb]{0.82, 0.82, 0.82} 57.08 
         & \cellcolor[rgb]{0.86, 0.86, 0.86} 45.59 
         & \cellcolor[rgb]{0.98, 0.98, 0.98} \phantom{0}7.75 
         & \cellcolor[rgb]{1.00, 1.00, 1.00} \phantom{0}0.00 
         & \cellcolor[rgb]{0.81, 0.81, 0.81} 57.21 \\ 
         & Quantization & \cellcolor[rgb]{0.94, 0.94, 0.94} 26.15 
         & \cellcolor[rgb]{0.99, 0.99, 0.99} \phantom{0}3.31 
         & \cellcolor[rgb]{0.98, 0.98, 0.98} \phantom{0}8.66 
         & \cellcolor[rgb]{0.95, 0.95, 0.95} 19.59 
         & \cellcolor[rgb]{0.91, 0.91, 0.91} 33.19 
         & \cellcolor[rgb]{0.99, 0.99, 0.99} \phantom{0}4.59 
         & \cellcolor[rgb]{0.97, 0.97, 0.97} 10.80 
         & \cellcolor[rgb]{0.95, 0.95, 0.95} 20.92 \\ 
         & Compressor & \cellcolor[rgb]{0.98, 0.98, 0.98} \phantom{0}9.30 
         & \cellcolor[rgb]{1.00, 1.00, 1.00} \phantom{0}1.02 
         & \cellcolor[rgb]{1.00, 1.00, 1.00} \phantom{0}0.00 
         & \cellcolor[rgb]{1.00, 1.00, 1.00} \phantom{0}0.00 
         & \cellcolor[rgb]{0.96, 0.96, 0.96} 14.63 
         & \cellcolor[rgb]{0.99, 0.99, 0.99} \phantom{0}4.14 
         & \cellcolor[rgb]{1.00, 1.00, 1.00} \phantom{0}0.00 
         & \cellcolor[rgb]{1.00, 1.00, 1.00} \phantom{0}0.00 \\ 
         & Opus & \cellcolor[rgb]{0.90, 0.90, 0.90} 36.27 
         & \cellcolor[rgb]{0.93, 0.93, 0.93} 27.55 
         & \cellcolor[rgb]{0.96, 0.96, 0.96} 17.35 
         & \cellcolor[rgb]{0.86, 0.86, 0.86} 47.38 
         & \cellcolor[rgb]{0.88, 0.88, 0.88} 40.22 
         & \cellcolor[rgb]{0.92, 0.92, 0.92} 30.58 
         & \cellcolor[rgb]{0.96, 0.96, 0.96} 17.32 
         & \cellcolor[rgb]{0.86, 0.86, 0.86} 46.17 \\ 
         \cmidrule{2-10}
         & Clipping & \cellcolor[rgb]{1.00, 1.00, 1.00} \phantom{0}1.22 
         & \cellcolor[rgb]{1.00, 1.00, 1.00} \phantom{0}0.23 
         & \cellcolor[rgb]{1.00, 1.00, 1.00} \phantom{0}0.00 
         & \cellcolor[rgb]{1.00, 1.00, 1.00} \phantom{0}0.00 
         & \cellcolor[rgb]{0.99, 0.99, 0.99} \phantom{0}6.62 
         & \cellcolor[rgb]{1.00, 1.00, 1.00} \phantom{0}0.92 
         & \cellcolor[rgb]{1.00, 1.00, 1.00} \phantom{0}0.00 
         & \cellcolor[rgb]{1.00, 1.00, 1.00} \phantom{0}0.00 \\ 
         & Overdrive & \cellcolor[rgb]{0.96, 0.96, 0.96} 15.30 
         & \cellcolor[rgb]{0.99, 0.99, 0.99} \phantom{0}6.19 
         & \cellcolor[rgb]{1.00, 1.00, 1.00} \phantom{0}0.11 
         & \cellcolor[rgb]{1.00, 1.00, 1.00} \phantom{0}0.00 
         & \cellcolor[rgb]{0.95, 0.95, 0.95} 21.04 
         & \cellcolor[rgb]{0.98, 0.98, 0.98} \phantom{0}8.53 
         & \cellcolor[rgb]{1.00, 1.00, 1.00} \phantom{0}0.05 
         & \cellcolor[rgb]{1.00, 1.00, 1.00} \phantom{0}0.00 \\ 
         & Equalizer & \cellcolor[rgb]{1.00, 1.00, 1.00} \phantom{0}1.75 
         & \cellcolor[rgb]{1.00, 1.00, 1.00} \phantom{0}0.23 
         & \cellcolor[rgb]{1.00, 1.00, 1.00} \phantom{0}0.00 
         & \cellcolor[rgb]{1.00, 1.00, 1.00} \phantom{0}0.03 
         & \cellcolor[rgb]{0.98, 0.98, 0.98} \phantom{0}9.56 
         & \cellcolor[rgb]{1.00, 1.00, 1.00} \phantom{0}0.90 
         & \cellcolor[rgb]{1.00, 1.00, 1.00} \phantom{0}0.00 
         & \cellcolor[rgb]{1.00, 1.00, 1.00} \phantom{0}0.05 \\ 
         & Frequency masking & \cellcolor[rgb]{0.87, 0.87, 0.87} 43.32 
         & \cellcolor[rgb]{0.91, 0.91, 0.91} 33.11 
         & \cellcolor[rgb]{0.99, 0.99, 0.99} \phantom{0}2.94 
         & \cellcolor[rgb]{0.94, 0.94, 0.94} 24.40 
         & \cellcolor[rgb]{0.85, 0.85, 0.85} 49.99 
         & \cellcolor[rgb]{0.90, 0.90, 0.90} 36.55 
         & \cellcolor[rgb]{0.99, 0.99, 0.99} \phantom{0}4.66 
         & \cellcolor[rgb]{0.94, 0.94, 0.94} 23.89 \\ 
         & Noise gate & \cellcolor[rgb]{0.98, 0.98, 0.98} 10.56 
         & \cellcolor[rgb]{0.99, 0.99, 0.99} \phantom{0}2.56 
         & \cellcolor[rgb]{1.00, 1.00, 1.00} \phantom{0}0.13 
         & \cellcolor[rgb]{0.99, 0.99, 0.99} \phantom{0}2.56 
         & \cellcolor[rgb]{0.96, 0.96, 0.96} 18.34 
         & \cellcolor[rgb]{0.99, 0.99, 0.99} \phantom{0}4.21 
         & \cellcolor[rgb]{1.00, 1.00, 1.00} \phantom{0}0.26 
         & \cellcolor[rgb]{0.99, 0.99, 0.99} \phantom{0}3.12 \\ 
         & Noise reduction & \cellcolor[rgb]{0.96, 0.96, 0.96} 17.18 
         & \cellcolor[rgb]{0.97, 0.97, 0.97} 11.61 
         & \cellcolor[rgb]{1.00, 1.00, 1.00} \phantom{0}0.00 
         & \cellcolor[rgb]{1.00, 1.00, 1.00} \phantom{0}0.05 
         & \cellcolor[rgb]{0.94, 0.94, 0.94} 24.61 
         & \cellcolor[rgb]{0.97, 0.97, 0.97} 14.16 
         & \cellcolor[rgb]{1.00, 1.00, 1.00} \phantom{0}0.00 
         & \cellcolor[rgb]{1.00, 1.00, 1.00} \phantom{0}0.08 \\
         \midrule
         % & Avg. (Partially seen) & \cellcolor[rgb]{0.92, 0.92, 0.92} 31.63
         % & \cellcolor[rgb]{0.95, 0.95, 0.95} 19.84 
         % & \cellcolor[rgb]{0.95, 0.95, 0.95} 20.06 
         % & \cellcolor[rgb]{0.87, 0.87, 0.87} 43.17 
         % & \cellcolor[rgb]{0.90, 0.90, 0.90} 35.73 
         % & \cellcolor[rgb]{0.95, 0.95, 0.95} 21.40 
         % & \cellcolor[rgb]{0.95, 0.95, 0.95} 19.70 
         % & \cellcolor[rgb]{0.87, 0.87, 0.87} 44.08 \\ 
         % & Avg. (Unseen) & \cellcolor[rgb]{0.95, 0.95, 0.95} 19.49 
         % & \cellcolor[rgb]{0.98, 0.98, 0.98} 8.36 
         % & \cellcolor[rgb]{0.99, 0.99, 0.99} 2.77 
         % & \cellcolor[rgb]{0.97, 0.97, 0.97} 14.00 
         % & \cellcolor[rgb]{0.94, 0.94, 0.94} 26.37 
         % & \cellcolor[rgb]{0.98, 0.98, 0.98} 10.48 
         % & \cellcolor[rgb]{0.99, 0.99, 0.99} 3.19 
         % & \cellcolor[rgb]{0.97, 0.97, 0.97} 14.06 \\ 
         \multicolumn{2}{c}{Average w/o None} & \cellcolor[rgb]{0.94, 0.94, 0.94} 23.77 
         & \cellcolor[rgb]{0.97, 0.97, 0.97} 12.41 
         & \cellcolor[rgb]{0.98, 0.98, 0.98} \phantom{0}8.87 
         & \cellcolor[rgb]{0.94, 0.94, 0.94} 24.29 
         & \cellcolor[rgb]{0.92, 0.92, 0.92} 29.67 
         & \cellcolor[rgb]{0.97, 0.97, 0.97} 14.34 
         & \cellcolor[rgb]{0.98, 0.98, 0.98} \phantom{0}9.02 
         & \cellcolor[rgb]{0.94, 0.94, 0.94} 24.66 \\
        \bottomrule
    \end{tabular}
    }
    \label{tab:performance}
\end{table*}

%% file: template.bbl
\begin{thebibliography}{10}
\providecommand{\url}[1]{#1}
\csname url@samestyle\endcsname
\providecommand{\newblock}{\relax}
\providecommand{\bibinfo}[2]{#2}
\providecommand{\BIBentrySTDinterwordspacing}{\spaceskip=0pt\relax}
\providecommand{\BIBentryALTinterwordstretchfactor}{4}
\providecommand{\BIBentryALTinterwordspacing}{\spaceskip=\fontdimen2\font plus
\BIBentryALTinterwordstretchfactor\fontdimen3\font minus \fontdimen4\font\relax}
\providecommand{\BIBforeignlanguage}[2]{{%
\expandafter\ifx\csname l@#1\endcsname\relax
\typeout{** WARNING: IEEEtran.bst: No hyphenation pattern has been}%
\typeout{** loaded for the language `#1'. Using the pattern for}%
\typeout{** the default language instead.}%
\else
\language=\csname l@#1\endcsname
\fi
#2}}
\providecommand{\BIBdecl}{\relax}
\BIBdecl

\bibitem{jung2022_aasist}
J.-w. Jung, H.-S. Heo, H.~Tak, H.-j. Shim \emph{et~al.}, ``{AASIST}: Audio anti-spoofing using integrated spectro-temporal graph attention networks,'' in \emph{Proc. {ICASSP}}, 2022, pp. 6367--6371.

\bibitem{zhu2024_slim}
Y.~Zhu, S.~Koppisetti, T.~Tran, and G.~Bharaj, ``{SLIM}: Style-linguistics mismatch model for generalized audio deepfake detection,'' in \emph{Proc. {NeurIPS}}, 2024.

\bibitem{wang2020_asvspoof2019}
X.~Wang, J.~Yamagishi, M.~Todisco, and {Others}, ``{ASVspoof} 2019: A large-scale public database of synthesized, converted and replayed speech,'' \emph{Computer Speech \& Language}, vol.~64, p. 101114, 2020.

\bibitem{xie2025_codecfake}
Y.~Xie, Y.~Lu, R.~Fu, Z.~Wen \emph{et~al.}, ``The codecfake dataset and countermeasures for the universally detection of deepfake audio,'' \emph{{IEEE} Trans. Audio, Speech, Lang. Process.}, vol.~33, pp. 386--400, 2025.

\bibitem{liu2024_timbre}
C.~Liu, J.~Zhang, T.~Zhang, X.~Yang \emph{et~al.}, ``{Detecting voice cloning attacks via Timbre watermarking},'' in \emph{Proc. Netw. Distrib. Syst. Secur. Symp.}, 2024.

\bibitem{roman2024_audioseal}
R.~S. Roman, P.~Fernandez, H.~Elsahar, A.~Défossez \emph{et~al.}, ``Proactive detection of voice cloning with localized watermarking,'' in \emph{Proc. {ICML}}, 2024.

\bibitem{chen2023_wavmark}
G.~Chen, Y.~Wu, S.~Liu, T.~Liu \emph{et~al.}, ``{WavMark}: Watermarking for audio generation,'' \emph{arXiv preprint arXiv:2308.12770}, 2023.

\bibitem{liu2024_audiomarkbench}
H.~Liu, M.~Guo, Z.~Jiang, L.~Wang \emph{et~al.}, ``{AudioMarkBench}: Benchmarking robustness of audio watermarking,'' in \emph{Proc. {NeurIPS} Datasets and Benchmarks Track}, 2024.

\bibitem{zhang2021_oneclass}
Y.~Zhang, F.~Jiang, and Z.~Duan, ``One-class learning towards synthetic voice spoofing detection,'' \emph{{IEEE} Signal Processing Letters}, vol.~28, pp. 937--941, 2021.

\bibitem{de2023encodec}
A.~Défossez, J.~Copet, G.~Synnaeve, and Y.~Adi, ``{High fidelity neural audio compression},'' \emph{Transactions on Machine Learning Research}, 2023.

\bibitem{kinnunen18b_dcf}
T.~Kinnunen, K.~A. Lee, H.~Delgado, N.~Evans \emph{et~al.}, ``t-{DCF}: A detection cost function for the tandem assessment of spoofing countermeasures and automatic speaker verification,'' in \emph{Proc. Odyssey}, 2018, pp. 312--319.

\bibitem{kawa23_defense}
P.~Kawa, M.~Plata, and P.~Syga, ``Defense against adversarial attacks on audio {deepFake} detection,'' in \emph{Proc. Interspeech}, 2023, pp. 5276--5280.

\bibitem{tak2022_rawboost}
H.~Tak, M.~Kamble, J.~Patino, M.~Todisco \emph{et~al.}, ``{RawBoost}: A raw data boosting and augmentation method applied to automatic speaker verification anti-spoofing,'' in \emph{Proc. {ICASSP}}, 2022, pp. 6382--6386.

\bibitem{iso/iecjtc1/sc37ISO2023}
\BIBentryALTinterwordspacing
{International Organization for Standardization}, \emph{{Information Technology — Biometric Presentation Attack Detection — Part 1: Framework}}, International Organization for Standardization Std. ISO/IEC 30\,107-1:2023, 2023. [Online]. Available: \url{https://www.iso.org/standard/83828.html}
\BIBentrySTDinterwordspacing

\bibitem{bishopPattern2006}
C.~M. Bishop, \emph{Pattern Recognition and Machine Learning}, ser. Information Science and Statistics.\hskip 1em plus 0.5em minus 0.4em\relax New York: Springer, 2006.

\bibitem{wu2015spoofing}
Z.~Wu, N.~Evans, T.~Kinnunen, J.~Yamagishi \emph{et~al.}, ``Spoofing and countermeasures for speaker verification: A survey,'' \emph{Speech Communication}, vol.~66, pp. 130--153, 2015.

\bibitem{todisco2017constant}
M.~Todisco, H.~Delgado, and N.~Evans, ``{Constant Q cepstral coefficients: A spoofing countermeasure for automatic speaker verification},'' \emph{Computer Speech \& Language}, vol.~45, pp. 516--535, 2017.

\bibitem{kinnunen2017asvspoof}
T.~Kinnunen, N.~Evans, J.~Yamagishi, K.~A. Lee \emph{et~al.}, ``{ASVspoof 2017: Automatic speaker verification spoofing and countermeasures challenge evaluation plan},'' in \emph{Proc. Interspeech}, 2017, pp. 2--6.

\bibitem{liu_asvspoof_2023}
X.~Liu, X.~Wang, M.~Sahidullah, J.~Patino \emph{et~al.}, ``{ASVspoof} 2021: Towards spoofed and deepfake speech detection in the wild,'' \emph{{IEEE} Trans. Audio, Speech, Lang. Process.}, vol.~31, pp. 2507--2522, 2023.

\bibitem{tak22_w2v_aasist}
H.~Tak, M.~Todisco, X.~Wang, J.-w. Jung \emph{et~al.}, ``Automatic speaker verification spoofing and deepfake detection using wav2vec 2.0 and data augmentation,'' in \emph{Proc. Odyssey}, 2022, pp. 112--119.

\bibitem{wang2023_copy_synthesis}
X.~Wang and J.~Yamagishi, ``Spoofed training data for speech spoofing countermeasure can be efficiently created using neural vocoders,'' in \emph{Proc. {ICASSP}}, 2023, pp. 1--5.

\bibitem{guragain2024_speech}
A.~Guragain, T.~Liu, Z.~Pan, H.~B. Sailor \emph{et~al.}, ``Speech foundation model ensembles for the controlled singing voice deepfake detection ({CtrSVDD}) challenge 2024,'' in \emph{Proc. {SLT}}, 2024, pp. 774--781.

\bibitem{ji2024_speechwatermark}
S.~Ji, Z.~Jiang, J.~Zuo, M.~Fang \emph{et~al.}, ``Speech watermarking with discrete intermediate representations,'' in \emph{Proc. {AAAI}}, 2025.

\bibitem{singh24_silentcipher}
M.~K. Singh, N.~Takahashi, W.~Liao, and Y.~Mitsufuji, ``{SilentCipher}: Deep audio watermarking,'' in \emph{Proc. Interspeech}, 2024, pp. 2235--2239.

\bibitem{cheng2024hifi}
X.~Cheng, Y.~Wang, C.~Liu, D.~Hu \emph{et~al.}, ``{HiFi}-{GANw}: Watermarked speech synthesis via fine-tuning of {HiFi}-{GAN},'' \emph{{IEEE} Signal Processing Letters}, vol.~31, pp. 2440--2444, 2024.

\bibitem{wuAdversarial2023}
S.~Wu, J.~Liu, Y.~Huang, H.~Guan \emph{et~al.}, ``Adversarial audio watermarking: Embedding watermark into deep feature,'' in \emph{Proc. {ICME}}, 2023, pp. 61--66.

\bibitem{o2024maskmark}
P.~O'Reilly, Z.~Jin, J.~Su, and B.~Pardo, ``{MaskMark}: Robust neuralwatermarking for real and synthetic speech,'' in \emph{Proc. {ICASSP}}, 2024, pp. 4650--4654.

\bibitem{van2007introduction}
D.~A. Van~Leeuwen and N.~Br{\"u}mmer, \emph{An introduction to application-independent evaluation of speaker recognition systems}.\hskip 1em plus 0.5em minus 0.4em\relax Springer, 2007.

\bibitem{brummer2021out}
N.~Brümmer, L.~Ferrer, and A.~Swart, ``Out of a hundred trials, how many errors does your speaker verifier make?'' in \emph{Proc. Interspeech}, 2021, pp. 1059--1063.

\bibitem{snyder_musan_2015}
D.~Snyder, G.~Chen, and D.~Povey, ``{MUSAN: A music, speech, and noise corpus},'' \emph{arXiv preprint arXiv:1510.08484}, 2015.

\bibitem{ko_study_2017}
T.~Ko, V.~Peddinti, D.~Povey, M.~L. Seltzer \emph{et~al.}, ``A study on data augmentation of reverberant speech for robust speech recognition,'' in \emph{Proc. {ICASSP}}, 2017, pp. 5220--5224.

\bibitem{sobot_peter_2023_7817838}
\BIBentryALTinterwordspacing
P.~Sobot, ``Pedalboard,'' Jul. 2021. [Online]. Available: \url{https://doi.org/10.5281/zenodo.7817838}
\BIBentrySTDinterwordspacing

\bibitem{valin2012definition}
J.-M. Valin, K.~Vos, and T.~Terriberry, ``Definition of the opus audio codec,'' Tech. Rep., 2012.

\bibitem{kumar2023high}
R.~Kumar, P.~Seetharaman, A.~Luebs, I.~Kumar \emph{et~al.}, ``High-fidelity audio compression with improved {RVQGAN},'' in \emph{Proc. {NeurIPS}}, vol.~36, 2023, pp. 27\,980--27\,993.

\bibitem{ji2024wavtokenizer}
S.~Ji, Z.~Jiang, W.~Wang, Y.~Chen \emph{et~al.}, ``{WavTokenizer}: An efficient acoustic discrete codec tokenizer for audio language modeling,'' in \emph{Proc. {ICLR}}, 2025.

\bibitem{paszke2019pytorch}
A.~Paszke, S.~Gross, F.~Massa, A.~Lerer \emph{et~al.}, ``{PyTorch: An imperative style, high-performance deep learning library},'' in \emph{Proc. {NeurIPS}}, 2019.

\bibitem{sainburg2020finding}
T.~Sainburg, M.~Thielk, and T.~Q. Gentner, ``Finding, visualizing, and quantifying latent structure across diverse animal vocal repertoires,'' \emph{PLoS Computational Biology}, vol.~16, no.~10, p. e1008228, 2020.

\bibitem{tim_sainburg_2019_3243139}
\BIBentryALTinterwordspacing
T.~Sainburg, ``timsainb/noisereduce: v1.0,'' Jun. 2019. [Online]. Available: \url{https://doi.org/10.5281/zenodo.3243139}
\BIBentrySTDinterwordspacing

\bibitem{schroeter2022deepfilternet}
H.~Schröter, A.~N. Escalante-B., T.~Rosenkranz, and A.~Maier, ``{DeepFilterNet}: A low complexity speech enhancement framework for full-band audio based on deep filtering,'' in \emph{Proc. {ICASSP}}, 2022, pp. 7407--7411.

\bibitem{schroeter2023deepfilternet3}
H.~Schröter, T.~Rosenkranz, A.~N. Escalante-B., and A.~Maier, ``{DeepFilterNet}: Perceptually motivated real-time speech enhancement,'' in \emph{Proc. Interspeech}, 2023, pp. 2008--2009.

\bibitem{mcfee2015librosa}
B.~McFee, C.~Raffel, D.~Liang, D.~P. Ellis \emph{et~al.}, ``librosa: Audio and music signal analysis in python,'' in \emph{Proc. {SciPy}}, 2015, pp. 18--25.

\bibitem{baevski2020wav2vec}
A.~Baevski, Y.~Zhou, A.~Mohamed, and M.~Auli, ``wav2vec 2.0: A framework for self-supervised learning of speech representations,'' in \emph{Proc. {NeurIPS}}, 2020.

\end{thebibliography}
